\RequirePackage{fix-cm}
\documentclass[smallextended]{svjour3}       
\smartqed  
\usepackage{graphicx}
\usepackage{siunitx}
\usepackage{amssymb}
\usepackage[center]{caption}
\usepackage[colorlinks]{hyperref}
\usepackage{epsfig}
\begin{document}

\title{Channel Capacity Enhancement of SWIPT based CNOMA Downlink Transmission with Orbital Angular Momentum over Rician Fading Channel}

\subtitle{}

\titlerunning{Channel Capacity Enhancement of SWIPT based CNOMA Downlink Transmission ..}        

\author{Ahmed Al Amin \and Soo Young Shin \and Bhaskara Narottama  
}


\institute{A.A. Amin \at
              Department of IT Convergence Engineering, Kumoh National Institute of Technology, Gumi, South Korea \\
              \email{amin@kumoh.ac.kr}           
            \and
           S. Y. Shin \at
              Department of IT Convergence Engineering, Kumoh National Institute of Technology, Gumi, South Korea \\
              \email{wdragon@kumoh.ac.kr}
              \and
               B. Narottama \at
              Department of IT Convergence Engineering, Kumoh National Institute of Technology, Gumi, South Korea \\
              \email{bhaskaranarottama@gmail.co}
}

\date{}

\maketitle

\begin{abstract}
In this paper, orbital angular momentum (OAM) is integrated with power splitting based simultaneous
wireless information and power transfer (SWIPT-PS) and cooperative non-orthogonal multiple access (CNOMA-SWIPT-PS) downlink transmission, called as CNOMA-SWIPT-PS-OAM. The proposed CNOMA-SWIPT-PS-OAM scheme enhances the spectral efficiency and energy efficiency as well. The SWIPT-PS will energize the relaying operation from cell center user (CCU) to cell edge user (CEU). Different symbols are transmitting towards the users simultaneously from the base station by utilizing different OAM modes on the time slot of relaying (CCU to CEU) for CNOMA-SWIPT-PS. The CCU, CEU and sum capacities and energy efficiency of the proposed scheme over Rician fading channels are investigated in this paper. Finally, the effectiveness of the proposed schemes over existing schemes and conventional orthogonal multiple access based scheme are demonstrated through the result analysis.   
\end{abstract}

\section{Introduction}
Spectral efficiency along with energy efficiency (EE) are two vital challenges that need to be satisfied among the several requirements to achieve the upcoming challenges of the future wireless networks [1,2]. To deal with these challenges non-orthogonal multiple access (NOMA) and simultaneous wireless information and power transfer (SWIPT) can be integrated. Moreover, by utilizing the different modes of orbital angular momentum (OAM), additional symbols can be transferred to cell center user (CCU) and cell edge user (CEU) to enhance the channel capacities as well. These are the main focus of this paper. 

NOMA is an interesting and suitable solution to enhance spectral efficiency and serve a large number of users [3,4].  
In NOMA based system, the users are served simultaneously by the same code at the same frequency, time but different power levels. 
Successive Interference Cancellation (SIC) is used at the received end to separate the superimposed messages transmitted by the source [5].      

Cooperative relaying is an effective way to enhance the reliability and coverage area by utilizing CCU as a relay for the CEU or using a dedicated relay which assists the NOMA users [6-9]. 
In general, the user closer from the base station (BS) is called CCU and the user further from BS is called CEU. 
NOMA with cooperative relaying is known as cooperative NOMA (CNOMA) [7-9].

Furthermore, RF signals can be utilized to energize the energy-constrained nodes. [10-11]
This will enhance the battery lifetime by utilizing the SWIPT technique [12].
SWIPT, which can extract energy and information simultaneously from the ambient RF signals using power splitting or time switching receiver architecture, can provide RF energy harvesting (EH) at the energy-constrained nodes [13].  

In present research work, CNOMA and SWIPT are integrated together which is known as CNOMA-SWIPT to provide spectral efficiency and energy efficiency as well [14-15].
CNOMA-SWIPT protocol can be categorized into two different types. 
In the first type, the CCU which is an energy-constrained node is used as a relay for the CEU [15-18].
In the second type, energy harvesting is used for the dedicated relay which relaying information for all NOMA users. 
The first type of CNOMA-SWIPT is the main consideration of this paper since dedicated relay is not available in every cases [19-20]. 
In the previous works, the ergodic sum capacity and outage probabilities are analyzed for the CNOMA-SWIPT protocol [21-22].  
In the work of Kader et. al. [22], two novel cooperative spectrum sharing protocols, i.e., time switching (TS) and power splitting (PS), were investigated for CNOMA-SWIPT, where dedicated relay is used as an energy-constrained relay for the CEU. The ergodic sum capacity (ESC) and outage probabilities (OP) are analyzed for the CNOMA-SWIPT over Nakagami-m fading channel. 
In the work of Shah et. al. [23], a CNOMA-SWIPT protocol was proposed over Rayleigh fading channels. 
Whereas the source is communicating with the user with a backscattering node and an EH based decode and forwards (DF) relay technique. 
Moreover, in [24], a transmit antenna selection scheme based on CNOMA-hybrid SWIPT protocol was proposed over the Rayleigh fading channel as well. Furthermore, the source is communicating with two users simultaneously via the support of an EH based decode and forwards (DF) relay technique. Furthermore, R. Jiao et. al. analyze the capacity of CNOMA over the Rician fading channel. However, the capacity enhancement and empowering the relaying node to enhance the energy efficiency are not analyzed to fulfill the requirements of the upcoming cellular communication system [25].    

Considering all the above techniques, a suitable solution is required to energize the relay operation and provides enhanced channel capacity simultaneously without any additional resources. That is why CNOMA downlink transmission with PS based SWIPT protocol (CNOMA-SWIPT-PS) is a viable solution for simplicity and provides better capacities than TS as well [22]. So, CNOMA-SWIPT-PS is considered in this paper to energize the relay operation. OAM utilizes a new degree of freedom which is known as  OAM mode for signal transmission [26-27]. Moreover, OAM  exploits the phase variation with respect to the azimuth angle of the propagated electromagnetic waves. To enhance the capacity of CNOMA-SWIPT-PS, OAM is a potential candidate to transmit different symbols by utilizing different modes of OAM simultaneously to enhance the channel capacities without any interference [26-28]. That is why a novel scheme is proposed here by utilizing the time slot of relaying in the case of CNOMA-SWIPT-PS protocol. Two different additional symbols will be transmitted by different OAM modes towards the users from the base station (BS) by utilizing the same time slot for relaying from CCU to CEU. This proposed scheme can enhance the capacity of the CNOMA with PS SWIPT protocol without any interference and this scheme does not any required additional resources (e.g. time slot or frequency band) as well. Moreover, energy efficiency (EE) will be also enhanced by utilizing the proposed technique as well. The principal contribution of this paper is given below: 

\begin{itemize}
\item  PS based CNOMA-SWIPT-PS is considered over the Rician fading channel.
\item OAM is integrated with CNOMA-SWIPT-PS protocol to achieve higher channel capacities. The proposed scheme of OAM with CNOMA-SWIPT-PS is mentioned as CNOMA-SWIPT-PS-OAM in this paper.  
\item To analyze the capacity enhancement of the proposed CNOMA-SWIPT-PS-OAM scheme, capacity of CCU, the capacity of CEU, and sum capacity (SC) are analyzed and investigated. Moreover, the capacities of the proposed scheme are compared with CNOMA based conventional SWIPT protocols (CNOMA-SWIPT-PS and CNOMA-SWIPT-TS) and conventional orthogonal multiple access (OMA) based SWIPT-PS with OAM (OMA-SWIPT-PS-OAM) as well.
\item EE is another performance analysis metric that is analyzed here as well. Hence the EE comparison for the proposed scheme is compared with CNOMA and OMA based conventional SWIPT protocols as well. 
\item The impact of the different parameters over capacities and EE for the proposed scheme is analyzed and compared with CNOMA-SWIPT-PS, CNOMA-SWIPT-TS, and OMA-SWIPT-PS-OAM as well to analyze the effectiveness of the proposed scheme. 
\end{itemize}

The rest of this paper is organized as follows. Section 2 describes the proposed CNOMA-SWIPT-PS-OAM system model and the proposed protocol as well. The capaciti analysis and OMA based SWIPT-PS scheme are also presented in this section. Section 3 exhibits numerical results. This paper is concluded in Section 4.

\section{System Model}

A two-phase CNOMA-SWIPT-PS network with a BS which is considered as a source and two users (a CCU and a CEU) is considered. The BS directly communicates with the CCU called $UE_1$ and CEU called $UE_2$ as well. There is two line of sight (LOS) links that are considered from BS to $UE_1$ and $UE_2$ accordingly to send additional symbols directly to the users as well by different OAM modes. To enhance the data reliability at $UE_2$ and enhance the coverage area, $UE_1$ perform the energy harvested based decode and forward (DF) relaying from $UE_1$ to $UE_2$. The proposed network model is shown in Figure 1. Hereafter, subscript $s$, $1$ and $2$ denote BS, $UE_1$ and $UE_2$ respectively. Assume that the channel coefficient $h_{p,q}$ between any two nodes $p$ and $q$ $(p,q \epsilon \{s, UE_1, UE_2\}$ and $i \neq j$) is subjected to Rician fading plus Additive White Gaussian Noise (AWGN). Whereas, $\Omega_{s,1}$, $\Omega_{s,2}$, and $\Omega_{1,2}$ are the average power of BS to $UE_1$, BS to $UE_2$, and $UE_1$ to $UE_2$ relay link respectively, where $\Omega_{s,2}<\Omega_{s,1}$ [25]. The total transmits power of S for direct transmissions (NOMA and OAM) and power transmit from $UE_1$ for relaying are denoted by $P$ and $P_1$ respectively. Moreover, $l_{s,1}$ and $l_{s,2}$ are two different OAM modes as well to transmit additional symbols to $UE_1$ and $UE_2$ respectively from the BS to enhance the capacities. The data communication policy along with the signal-to-interference-plus-noise ratio (SINR) model of the proposed CNOMA-SWIPT-PS-OAM scheme are discussed explicitly in section 2.1 and 2.2, respectively.            

\begin{figure}[!t]
\centering
\includegraphics[width=0.9\textwidth]{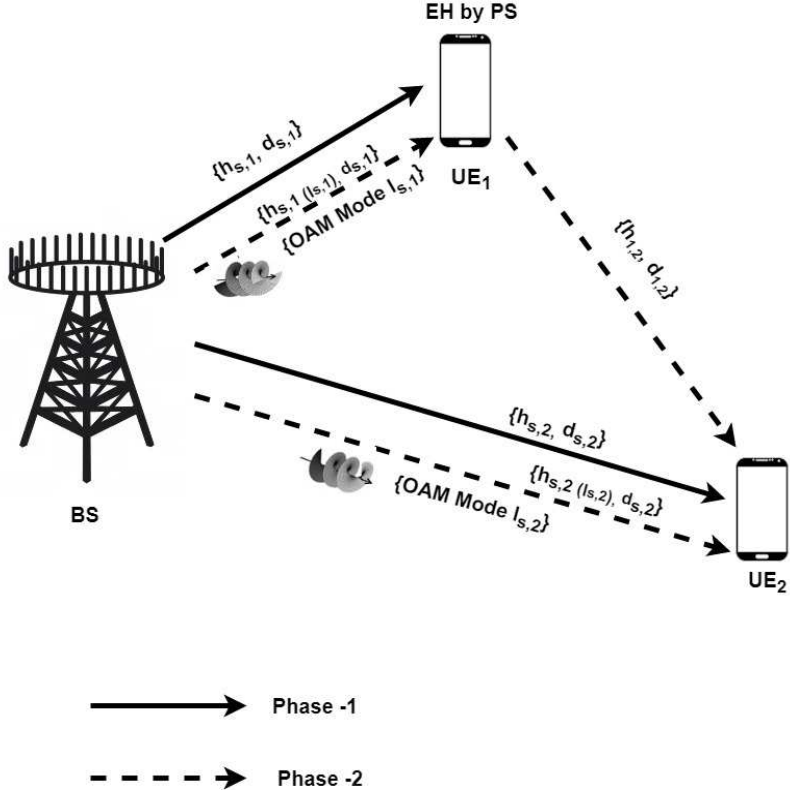}
\caption{Proposed system model for CNOMA-SWIPT-PS-OAM}
\label{image-myimage}
\end{figure}
\begin{figure}[!t]
\centering
\includegraphics[width=0.8\textwidth]{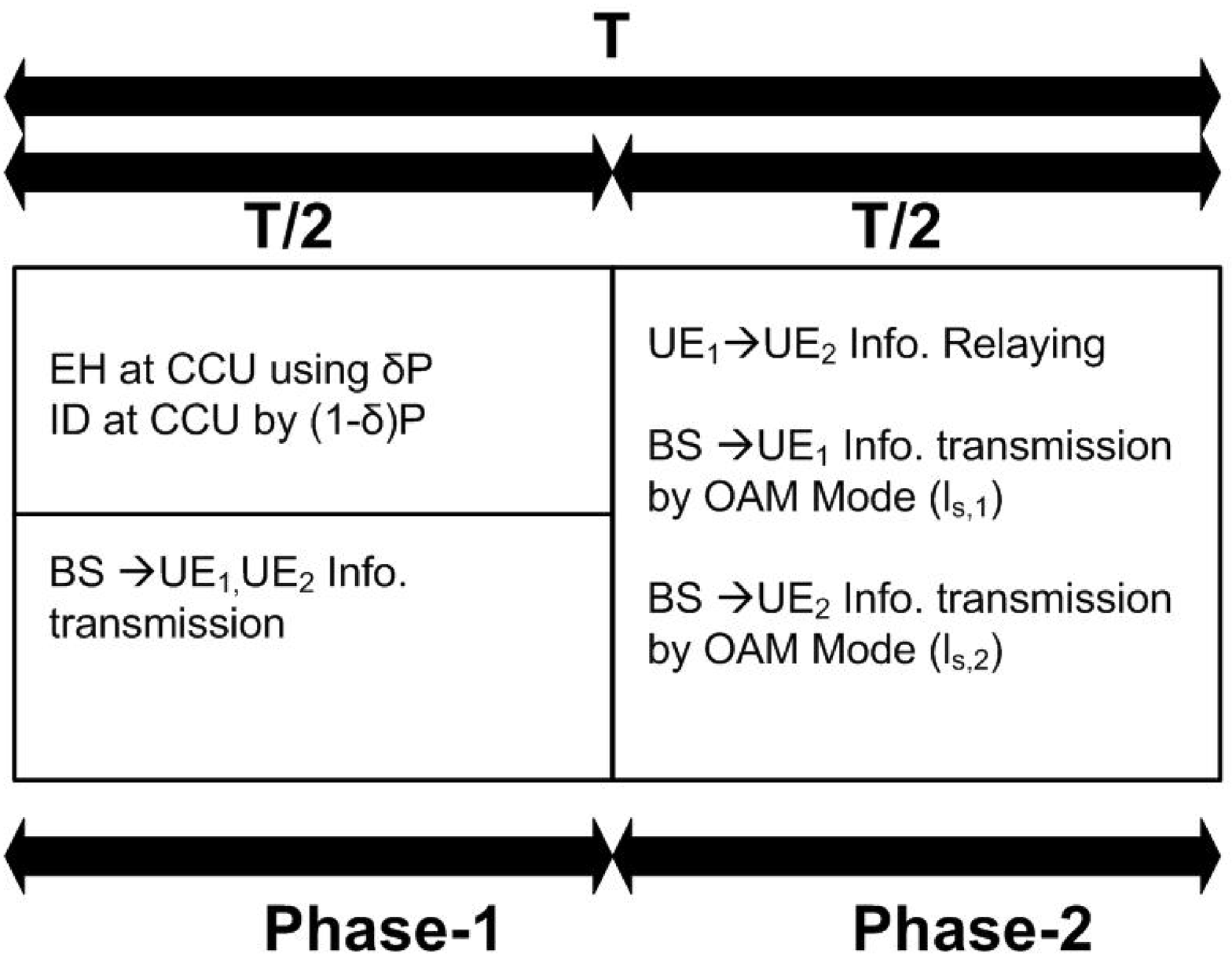}
\caption{Proposed protocol for CNOMA-SWIPT-PS-OAM}
\label{image-myimage}
\end{figure}

\subsection{CNOMA-SWIPT-PS-OAM}

In this strategy, it is considered that $UE_1$ acts as a PS based EH relay. Moreover, additional symbols are transmitted to the users as well by utilizing different OAM modes. The proposed system model and protocol summary are presented in Figure 1 and Figure 2 respectively. $T$ is the total duration of downlink transmission of the CNOMA-SWIPT-PS-OAM scheme which is shown in Figure 2.  

\subsubsection{Phase-1}

Following the principle of downlink NOMA, BS transmits a superimposed composite signal $A=\sqrt{p_{N}P}x_1+\sqrt{p_{F}P}x_2$ for a duration of $T/2$, where $x_1$, $x_2$ are the data symbols and $p_N$, $p_F$ are power allocating factors respectively. Note that $x_1$, $p_N$ and $x_2$, $p_F$ are assigned to $UE_1$ and $UE_2$, respectively. Moreover, the received power at $UE_1$ is split in the ratio of $\delta P: (1-\delta)P (0 \leq \delta \leq 1)$, where $\delta P$ and $(1-\delta)P$ are for EH and information decoding (ID) respectively [22]. The transmitted power $P_1$ of $UE_1$ by harvested energy for DF relaying is given below

\begin{equation}
P_{1}={\eta \delta P{{|h_{s,1}|}^2}}.
\end{equation}
Where, $\eta$ is the energy conversion efficiency, where $0 \leq \eta \leq 1$ [14,18,22,24]. The received signals at $UE_1$ and $UE_2$ are respectively given by,

\begin{equation}
y_1=(\sqrt{p_{N}(1-\delta)P}x_1+\sqrt{p_{F}(1-\delta)P}x_2)|h_{s,1}|^2 + n_1,
\end{equation}

\begin{equation}
y_2=(\sqrt{p_{N}(1-\delta)P}x_1+\sqrt{p_{F}(1-\delta)P}x_2)|h_{s,2}|^2 + n_2,
\end{equation}

where $n_1 \sim CN(0,\sigma^2)$ and $n_2 \sim CN(0,\sigma^2)$ are the complex AWGN at $UE_1$ and $UE_2$ respectively with zero mean and variance $\sigma^2$. According to the downlink NOMA protocol, $UE_1$ first decodes $x_2$ and then performs SIC to decode own symbol $x_1$ [22]. Thus, the received SINR at $UE_1$ for $x_2$ and $x_1$ are respectively given by following equations 

\begin{equation}
\gamma_{x_1}^{t_1}={(1-\delta)\rho{{|h_{s,1}|}^2}p_{N}},
\end{equation}
\begin{equation}
\gamma_{x_2 \rightarrow x_1}^{t_1}=\frac{(1-\delta)\rho{{|h_{s,1}|}^2}p_F}{(1-\delta)\rho{{|h_{s,1}|}^2}p_{N}+1},
\end{equation}

where $\rho \triangleq \frac{P}{\sigma^2}$ is the transmit signal-to-noise ratio (SNR) by BS and $\gamma_{x_2 \rightarrow x_1}^{t_1}$ denotes the required SINR to decode symbol $x_2$. Afterwards, $UE_2$ can obtain $x_2$ directly by treating $x_1$ as noise. Therefore the received SINR at $UE_2$ for $x_2$ can be derived by following equation

\begin{equation}
\gamma_{x_2}^{t_1}=\frac{\rho{{|h_{s,2}|}^2}(1-\delta)p_F}{\rho{{|h_{s,2}|}^2}(1-\delta)p_{N}+1}.
\end{equation}

Note that $p_N+p_F=1-\delta$ and $p_N<p_F$ [22].

\subsubsection{Phase-2}

This phase is consists of cooperative DF relaying and BS transmits additional symbols to the $UE_1$ and $UE_2$ by different OAM modes to the users. The cooperative DF relay transmission from $UE_1$ to $UE_2$ and additional symbol transmission directly by different OAM modes from BS are performed on the remaining $T-T/2=T/2$ duration. The cooperative relay is performed by $UE_1$. $UE_1$ transmits the decoded symbol $\hat{x_2}$ to $UE_2$ by utilizing the harvested energy. The received signal at $UE_2$ by DF relaying is expressed below

\begin{equation}
\hat{y_2}=\sqrt{P_1} \hat{x_2}|h_{1,2}|^2 + n_2.
\end{equation}

Therefore, the received SINR at $UE_2$ by DF relaying is obtained as below

\begin{equation}
\gamma_{x_2}^{t_2}={\rho\eta \delta {{|h_{s,1}|}^2}{{|h_{1,2}|}^2}}.
\end{equation}

Two additional symbols $x_3$ for $UE_1$ and $x_4$ for $UE_2$ by using different LOS channels to $UE_1$ and $UE_2$ respectively from BS. BS transmit signal $B=\sqrt{P}x_3$ to $UE_1$ and signal $D=\sqrt{P}x_4$ to CEU for a duration of $T/2$ simultaneously. To mitigate the interference issue and effective transmission two different OAM modes $l_{s,1}$ and $l_{s,2}$ are considered for $UE_1$ and $UE_2$ respectively. The LOS channel coefficient $h_{y,z}(l_{y,z})$ between any two nodes y and $(y,z \epsilon \{s, UE_1, UE_2\}$ and $y \neq z$) is subjected to LOS channel. The received signal at $UE_1$ and $UE_2$ from the BS by different OAM modes are given below

\begin{equation}
y_3=(\sqrt{P} x_3)|h_{s,1,(l_{s,1})}|^2 + n_1.
\end{equation}

\begin{equation}
y_4=(\sqrt{P} x_4)|h_{s,2,(l_{s,2})}|^2 + n_2.
\end{equation}

Accordingly, the received SINR for symbol $x_3$ at $UE_1$ can be expressed as [27-31]
\begin{equation}
\gamma_{x_3}^{t_2}={\rho }{{\mu_1}},
\end{equation}

where $\mu_1$ is the singular value of the channel response matrix $(h_{s,1,(l_{s,1})})$ of CNOMA-SWIPT-PS-OAM system [22,27,30]. OAM beam has divergence in its high-intensity region which caused attenuation [27-30]. By using Fresnel-zone-plate lenses antenna at BS, this issue can be mitigated without affecting the helical phase profile of OAM beam [27-30]. So, similarly the received SINR for symbol $x_4$ at $UE_2$ can be expressed as [29-31]
\begin{equation}
\gamma_{x_4}^{t_2}={\rho }{{\mu_2}},
\end{equation}

where similarly as before $\mu_2$ is the singular value of the channel response matrix $(h_{s,2,(l_{s,2})})$ of the proposed system [22,27,30].  

\subsection{Achievable capacity analysis}
By considering normalized total time duration ($T=1$). The equations for the capacities are derived in following segments. 

\subsubsection{Capacity of $UE_1$}
$x_1$ and $x_3$ are received by $UE_1$. So, the achievable capacity of $UE_1$ is obtained as below by (4) and (11),
\begin{equation}
C_{UE_1}=\frac{1}{2}\log_2(1+\gamma_{x_1}^{t_1})+\frac{1}{2} \log_2(1+\gamma_{x_3}^{t_2}).
\end{equation}

\subsubsection{Capacity of $UE_2$}

Using (5), (6), (8) and (12) the achievable capacity of $UE_2$ for $x_2$ and $x_4$ can be achieved by following equation,
\begin{equation}
C_{UE_2}=\frac{1}{2}\log_2(1+min(\gamma_{x_2 \rightarrow x_1}^{t_1},\gamma_{x_2}^{t_1},\gamma_{x_2}^{t_2}))+\frac{1}{2} \log_2(1+\gamma_{x_4}^{t_2}).
\end{equation}

\subsubsection{Sum Capacity}

So, the SC can be achieved by following equation [27]

\begin{equation}
C_\text{sum} = C_{UE_1}+C_{UE_2}.
\end{equation}
Whereas, $C_{UE_1}$ and $C_{UE_2}$ are the capacity of $UE_1$ and $UE_2$ for the proposed CNOMA-SWIPT-PS-OAM technique. 

\subsection{Energy Efficiency of CNOMA-SWIPT-PS-OAM}
Energy efficiency is the ratio of ergodic sum channel capacity ($C_\text{sum}$) and the total transmitted power by the BS for direct transmissions and transmission power from CCU for relaying. BS transmits $P$ amount of power for direct transmissions for NOMA from BS. For additional symbol transmission by different OAM modes to the users from BS, $P$ amount of power is also transmitted from BS as well. $UE_1$ transmit $P_1$ for DF relaying from $UE_1$ to $UE_2$. So the EE can be derived as below for the proposed CNOMA-SWIPT-PS-OAM [34-35],
\begin{equation}
EE=\frac{C_{sum}}{2P+P_{1}}.
\end{equation}

\subsection{OMA-SWIPT-PS-OAM}
For a fair comparison with the prposed CNOMA-SWIPT-PS-OAM scheme, OMA-SWIPT-PS-OAM scheme is also devised in this paper as benchmark. For the OMA case, time division multiple access (TDMA) is considered here. In this case, BS transmits information signal for $UE_1$ and $UE_2$ independently in different time slots with total transmit power $P$. The different time slots allocated for $UE_1$ and $UE_2$ for different symbols (e.g. $x_1$, $x_2$, $x_3$ and $x_4$) and DF relaying of $x_2$ are denoted as $t_1$, $t_2$, $t_3$, and $t_4$ respectively. Moreover, PS based SWIPT is considered at $UE_1$ [16,18,22]. Hence, $UE_1$ utilizes the fraction $\delta$ of the received power for energy harvesting and the remaining  $1-\delta$ fraction for ID [18,22,27]. So different transmission for $x_1$, $x_2$ and DF relaying of $x_2$ utilizes single time slot for each transmission. Such as $x_1$ transmitted from BS to $UE_1$ in $t_1$ along with PS SWIPT based energy harvesting at $UE_1$. $x_2$ is transmitting from BS to $UE_2$ in $t_2$. Furthermore, DF relaying transmission of $x_2$ by utilizing the harvested energy from $UE_1$ to $UE_2$ is perform in $t_3$. The transmission of $x_3$ and $x_4$ by different OAM modes ($l_{s,1}$ and $l_{s,2}$) are performed in $t_4$. Because without interference by using different modes of OAM, different symbols can be transmitted to the users in same time slot ($t_4$). Moreover, since $T=1$ is considered here, so total time slot is equally splitted for each time slot for OMA-SWIPT-PS-OAM technique. 
Hence, $t_1=t_2=t_3=t_4=\frac{1}{4}$ are considered here, which are the duration of each time slot. So, the achievable capacity of $UE_1$ and $UE_2$ can be achieved as below for the OMA-SWIPT-PS-OAM technique [5,27,33]

\begin{equation}
C_{UE_1}^{OMA}=\frac{1}{4}\log_2(1+\rho (1-\delta) |h_{s,1}|^2)+\frac{1}{4} \log_2(1+{\rho }{{\mu_1}}).
\end{equation}

\begin{equation}
C_{UE_2}^{OMA}=\frac{1}{4}\log_2(1+min(\rho (1-\delta) |h_{s,2}|^2,\eta \delta \rho |h_{s,1}|^2|h_{1,2}|^2))+\frac{1}{4} \log_2(1+{\rho }{{\mu_2}}).
\end{equation}

\begin{equation}
C_\text{sum}^\text{OMA} = C_{UE_1}^{OMA}+C_{UE_2}^{OMA}.
\end{equation}
 Moreover, $C_{UE_1}^{OMA}$ and $C_\text{sum}^\text{OMA}$ are the capacity of CCU and CEU for OMA-SWIPT-PS-OAM technique. Moreover, EE can be derived as below for the OMA-SWIPT-PS-OAM [35-36],
\begin{equation}
EE_{OMA}=\frac{C_\text{sum}^\text{OMA}}{3P+P_{1}}.
\end{equation}
So the above equation shows that EE is related to the $C_\text{sum}^\text{OMA}$, $P$ and $P_{1}$ as before. 

\section{Numerical Results}

In this section, results for the user capacities, SC and EE of the proposed protocol and techniques are examined and explained. The impact of changes in transmit SNR $\rho$, allowed power splitting factor $\delta$, distance $d_{s,1}$ between BS and $UE_1$ on the performance of the considered system are discussed. Collinear placement of all nodes (e.g. BS, $UE_1$, and $UE_2$) and normalized distances between any two nodes are considered, where $d_{s,1}=0.5$, $d_{s,2}=1$, and $d_{1,2}=1-d_{s,1}$. $(\Omega_{s,2}=9)< (\Omega_{s,1}=\Omega_{1,2}=36)$ are considered for simulation [25,27]. Furthermore, normalized $P=1$, normalized $T=1$, $l_{s,1}=2$ and $l_{s,2}=1$ are considered here for simulation purpose. For performance comparison, simulation results for CCU ($UE_1$) capacity, CEU ($UE_2$) capacity and SC of CNOMA-SWIPT-PS, CNOMA-SWIPT-TS, CNOMA-SWIPT-PS-OAM, and OMA-SWIPT-PS-OAM are also provided. Note that similar simulation parameters are considered for the proposed and other compared schemes for consistency. 

\subsection{User Capacities and Sum Capacity}
The impact of $\rho$, $d_{s,1}$ and $\delta$ on the capacity of CCU, CEU, and SC of the proposed system is shown in this part. All figures are plotted for the parameters $K_{s,1} = K_{1,2} = 5$, and $K_{s,2} = 2$ due to Rician fading channel [25,27].  
\par
CCU ($UE_1$) capacity behavior with respect to (w.r.t) transmit SNR $\rho$ is demonstrated in Figure 3 for the proposed CNOMA-SWIPT-PS-OAM scheme and compared with CNOMA-SWIPT-PS, CNOMA-SWIPT-TS, and OMA-SWIPT-PS-OAM schemes [22]. Parameters $p_N=0.4$, $p_F=0.6$, $\eta=0.7$, $\delta=0.3$, $d_{s,1}=0.5$, $d_{s,2}=1$, and $d_{1,2}=1-d_{s,1}$ are set during the simulation purpose. CCU capacity of all schemes increases linearly with an increase of $\rho$. CNOMA-SWIPT-PS-OAM exhibits far better performance than other schemes in case of CCU capacity because of additional symbol transmission from BS to CCU by an OAM mode ($l_{s,1}$). In the case of OMA-SWIPT-PS-OAM, CCU capacity is degraded than the proposed scheme due to using a dedicated time slot for each separate transmission from BS to CCU, excluding OAM based transmission.   
\begin{figure}[!t]
\centering
\includegraphics[width=0.8\textwidth]{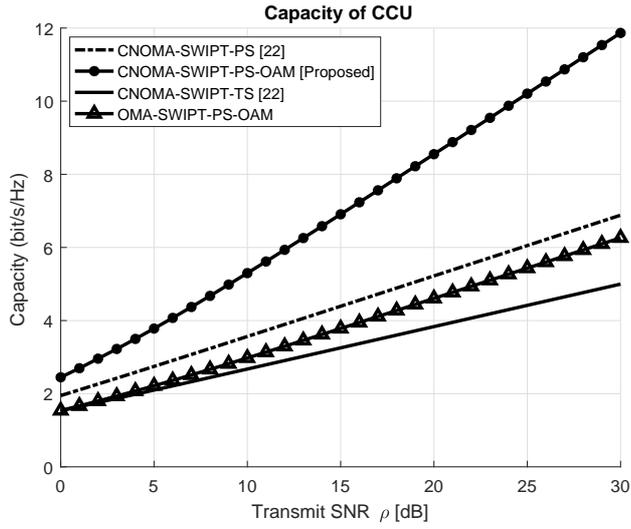}
\caption{Capacity comparisons of CCU with respect to transmit SNR $\rho$; $K_{s,1} = K_{1,2} = 5$, $K_{s,2} = 2$, $p_N=0.4$, $p_F=0.6$, $P=1$, $l_{s,1}=2$, $l_{s,2}=1$, $\delta=0.3$, $\eta=0.7$, $d_{s,1}=0.5$ and $d_{s,2}=1$.}
\label{image-myimage}
\end{figure}
\par
Figure 4 represents CEU ($UE_2$) capacity behavior w.r.t transmit SNR $\rho$ for the proposed CNOMA-SWIPT-PS-OAM and compared with CNOMA-SWIPT-PS, CNOMA-SWIPT-TS, and OMA-SWIPT-PS-OAM schemes as well. Parameters $p_N=0.4$, $p_F=0.6$, $\eta=0.7$, $\delta=0.3$, $d_{s,1}=0.5$, $d_{s,2}=1$, and $d_{1,2}=1-d_{s,1}$ are set during the simulation purpose as before. CEU capacity of all schemes increases linearly with increasing values of $\rho$. Figure 4 illustrates that in the case of CEU, the proposed scheme (CNOMA-SWIPT-PS-OAM) provides higher CEU capacity than other schemes as well. Similarly as before, an additional symbol is transmitting by utilizing the OAM beam with different OAM mode ($l_{s,2}$) on the time slot for relaying. Hence, the capacity is enhanced for the proposed technique at CEU. In the case of OMA-SWIPT-PS-OAM, CEU capacity is degraded than the proposed scheme due to using a dedicated time slot for each separate transmission from BS to CEU like as CCU, excluding OAM based transmission.  
\begin{figure}[!t]
\centering
\includegraphics[width=0.8\textwidth]{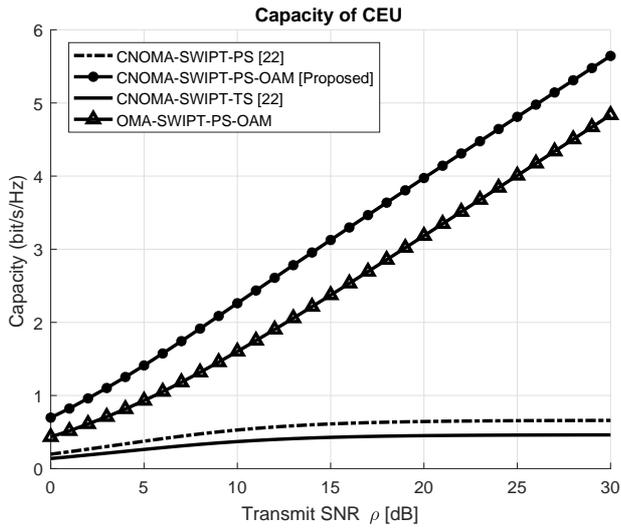}
\caption{Capacity comparisons of CEU with respect to transmit SNR $\rho$; $K_{s,1} = K_{1,2} = 5$, $K_{s,2} = 2$, $p_N=0.4$, $p_F=0.6$, $P=1$, $l_{s,1}=2$, $l_{s,2}=1$, $\delta=0.3$, $\eta=0.7$, $d_{s,1}=0.5$ and $d_{s,2}=1$.}
\label{image-myimage}
\end{figure}
\par
Figure 5 shows SC behavior w.r.t transmit SNR $\rho$ for the proposed CNOMA-SWIPT-PS-OAM and compared with CNOMA-SWIPT-PS, CNOMA-SWIPT-TS, and OMA-SWIPT-PS-OAM schemes as well. Parameters $p_N=0.4$, $p_F=0.6$, $\eta=0.7$, $\delta=0.3$, $d_{s,1}=0.5$, $d_{s,2}=1$, and $d_{1,2}=1-d_{s,1}$ are set during the simulation purpose as before. Figure 5 illustrates that SC of all schemes increases linearly with the increasing values of $\rho$. Moreover, the proposed technique (CNOMA-SWIPT-PS-OAM) provides higher SC than other conventional techniques as well. The higher SC is achieved by utilizing the different OAM modes to transmitting additional symbols to the CCU and CEU from the BS. Hence, the capacity of CCU and CEU for the proposed scheme are significantly higher than other existing schemes which is shown in Figure 3 and Figure 4. As a result, the SC of the proposed system is also enhanced significantly. In the case of OMA-SWIPT-PS-OAM, SC is lower than the proposed scheme because the capacities of CCU and CEU are comparatively lower than the proposed scheme (Which are shown in Figure 3 and Figure 4).    
\begin{figure}[!t]
\centering
\includegraphics[width=0.8\textwidth]{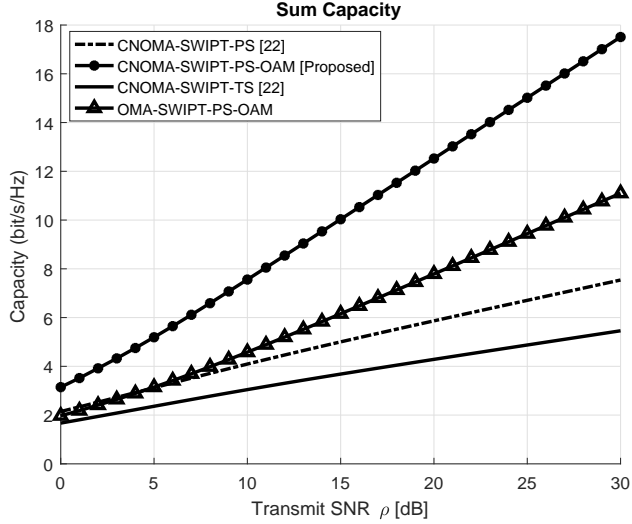}
\caption{SC comparisons with respect to transmit SNR $\rho$; $K_{s,1} = K_{1,2} = 5$, $K_{s,2} = 2$, $p_N=0.4$, $p_F=0.6$, $P=1$, $l_{s,1}=2$, $l_{s,2}=1$, $\delta=0.3$, $\eta=0.7$, $d_{s,1}=0.5$ and $d_{s,2}=1$.}
\label{image-myimage}
\end{figure}
\par
In Figure 6, the SC w.r.t $d_{s,1}$ is plotted for CNOMA-SWIPT-PS-OAM and compared with CNOMA-SWIPT-PS, CNOMA-SWIPT-TS, and OMA-SWIPT-PS-OAM schemes as well. Parameters $p_N=0.4$, $p_F=0.6$, $\eta=0.7$, $\delta=0.3$, and $\rho=15dB$ are set during the simulation purpose. Figure 6 shows that due to increasing values of $d_{s,1}$, the SC is decreasing for all cases. This is happening because due to increasing values of $d_{s,1}$, the channel condition between BS and CCU is degraded. Moreover, $p_N$ is lower than $p_F$ as well due to NOMA [22]. So, CCU cannot decode the information properly due to the degradation of the channel from BS to CCU. Figure 6 also illustrates that due to the increase of $d_{s,1}$, the SC of the proposed scheme is comparatively higher than other conventional schemes. Due to the increasing values of $d_{s,1}$, the channel condition between BS to CCU is degrading as well. But due to the simultaneous transmission of additional symbols from BS to the users by different OAM modes is performed on the time duration of relaying. So, the achieved SC of the proposed scheme is significantly higher than other techniques due to the increasing values of $d_{s,1}$. Due to utilizing the separate time slot for each transmission, except OAM based transmission. So, the OMA-SWIPT-PS-OAM scheme provides comparatively lower SC w.r.t $d_{s,1}$ than the proposed scheme which is illustrated in Figure 6 as well. 
\begin{figure}[!t]
\centering
\includegraphics[width=0.8\textwidth]{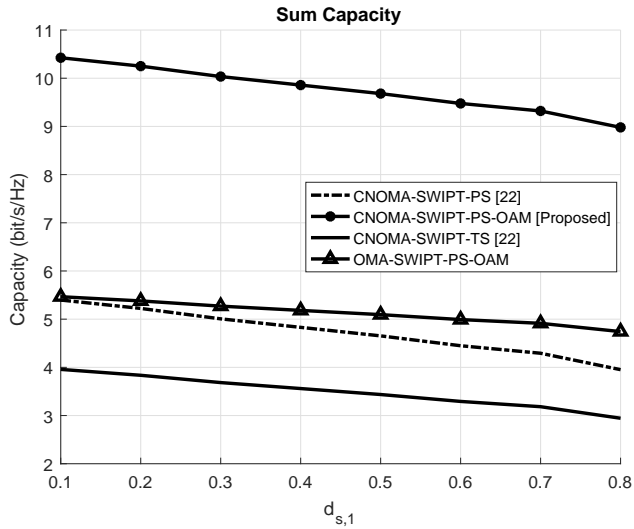}
\caption{SC comparisons with respect to $d_{s,1}$;$p_N=0.4$, $p_F=0.6$, $P=1$, $l_{s,1}=1$, $l_{s,2}=2$, $\delta=0.3$, $\eta=0.7$ and $\rho=15dB$.}
\label{image-myimage}
\end{figure}
\par
In Figure 7, the influence of $\delta$ over SC is plotted for CNOMA-SWIPT-PS-OAM and compared with CNOMA-SWIPT-PS, CNOMA-SWIPT-TS, and OMA-SWIPT-PS-OAM schemes as well. Parameters $p_N=0.4$, $p_F=0.6$, $\eta=0.7$, $d_{s,1}=0.5$,$d_{s,2}=1$, $d_{1,2}=1-d_{s,1}$, and $\rho=15dB$ are set during the simulation purpose. Figure 7 illustrates that due to increasing values of $\delta$, the SC is decreasing for all schemes except CNMOMA-SWIPT-TS. Because in case of TS based SWIPT, there is no impact of $\delta$ [22]. Because $\delta P$ is only used for energy harvesting for PS based SWIPT protocols. So, the higher values of $\delta P$ can enhance the harvested energy for relaying. In contrast, the amount of harvested energy ($(1-\delta)P$) for information decoding (ID) decreases as well for increasing values of $\delta$. Hence, the information cannot decode properly at CCU. As a result, the SC is decreased for the increasing values of $\delta$. Figure 7 shows that the proposed CNOMA-SWIPT-PS-OAM provides better SC than other schemes for different values of $\delta$. Because the additional information transmission from BS to the users by different OAM modes can enhance the SC of the proposed scheme than other compared schemes significantly. Due to utilizing the separate time slot for each transmission except OAM based transmission, the OMA-SWIPT-PS-OAM scheme provides comparatively lower SC w.r.t $\delta$ than the proposed scheme which is illustrated in Figure 7 as before.           
\begin{figure}[!t]
\centering
\includegraphics[width=0.8\textwidth]{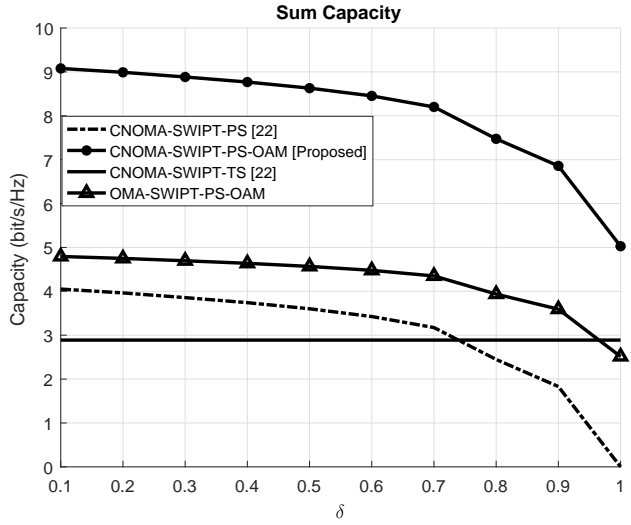}
\caption{SC comparisons with respect to $\delta$; $K_{s,1} = K_{s,2} = 5$, $K_{s,2} = 2$, $p_N=0.4$, $p_F=0.6$, $P=1$, $l_{s,1}=2$, $l_{s,2}=1$, $\eta=0.7$, $d_{s,1}=0.5$, $d_{s,2}=1$ and $\rho=15dB$.}
\label{image-myimage}
\end{figure}

\subsection{Energy Efficiency}
EE shows the ratio of SC and total transmitted power. Impact of $\rho$, $d_{s,1}$ and $\delta$ on EE of the proposed scheme is shown in this part. All figures are plotted for the parameters $K_{s,1} = K_{1,2} = 5$, and $K_{s,2} = 2$ due to Rician fading channel [25,27].

\par
EE w.r.t transmit SNR $\rho$ is demonstrated in Figure 8 for the proposed CNOMA-SWIPT-PS-OAM and compared with CNOMA-SWIPT-PS, CNOMA-SWIPT-TS, and OMA-SWIPT-PS-OAM schemes [22]. Parameters $p_N=0.4$, $p_F=0.6$, $\eta=0.7$, $\delta=0.3$, $d_{s,1}=0.5$, $d_{s,2}=1$, and $d_{1,2}=1-d_{s,1}$ are set during the simulation purpose as before. The EE is going downwards due to the increasing number of transmit SNR $\rho$ for every case. The transmitted power for relaying by the harvested energy is higher at CCU for higher values of $\rho$. That is why the EE is decreasing for higher $\rho$ in all cases. Moreover, the proposed CNOMA-SWIPT-PS-OAM scheme provides higher EE than other schemes due to higher SC can be achieved by different OAM mode based transmission from BS to the users. So, Figure 8 illustrates that the proposed technique provides significantly higher EE than other conventional techniques as well. Because the SC is higher for the proposed CNOMA-SWIPT-PS-OAM technique than other compared schemes which is shown in Figure 5. Moreover, the PS based SWIPT provides higher EE than TS based SWIPT for CNOMA as well which is also shown in Figure 8 as well. Due to utilizing the separate time slot for each transmission except OAM based transmission, the OMA-SWIPT-PS-OAM scheme provides comparatively lower EE than the proposed scheme w.r.t $\rho$ which is illustrated in Figure 8.     
\begin{figure}[!t]
\centering
\includegraphics[width=0.8\textwidth]{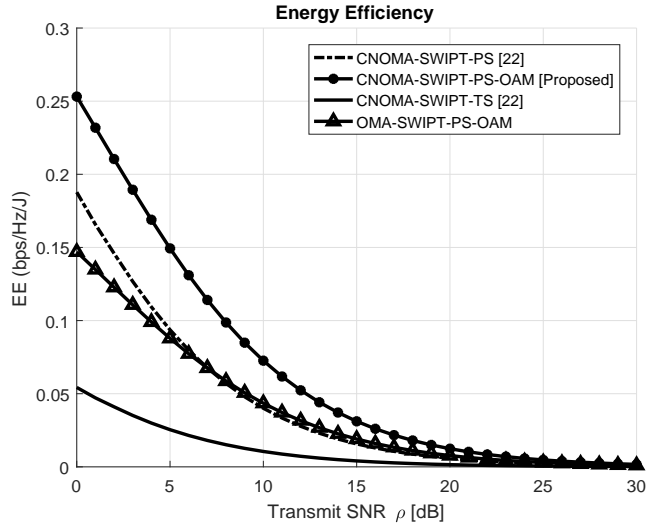}
\caption{EE comparisons with respect to transmit SNR $\rho$; $K_{s,1} = K_{1,2} = 5$, $K_{s,2} = 2$, $p_N=0.4$, $p_F=0.6$, $P=1$, $l_{s,1}=2$, $l_{s,2}=1$, $\delta=0.3$, $\eta=0.7$, $d_{s,1}=0.5$ and $d_{s,2}=1$.}
\label{image-myimage}
\end{figure}
\begin{figure}[!t]
\centering
\includegraphics[width=0.8\textwidth]{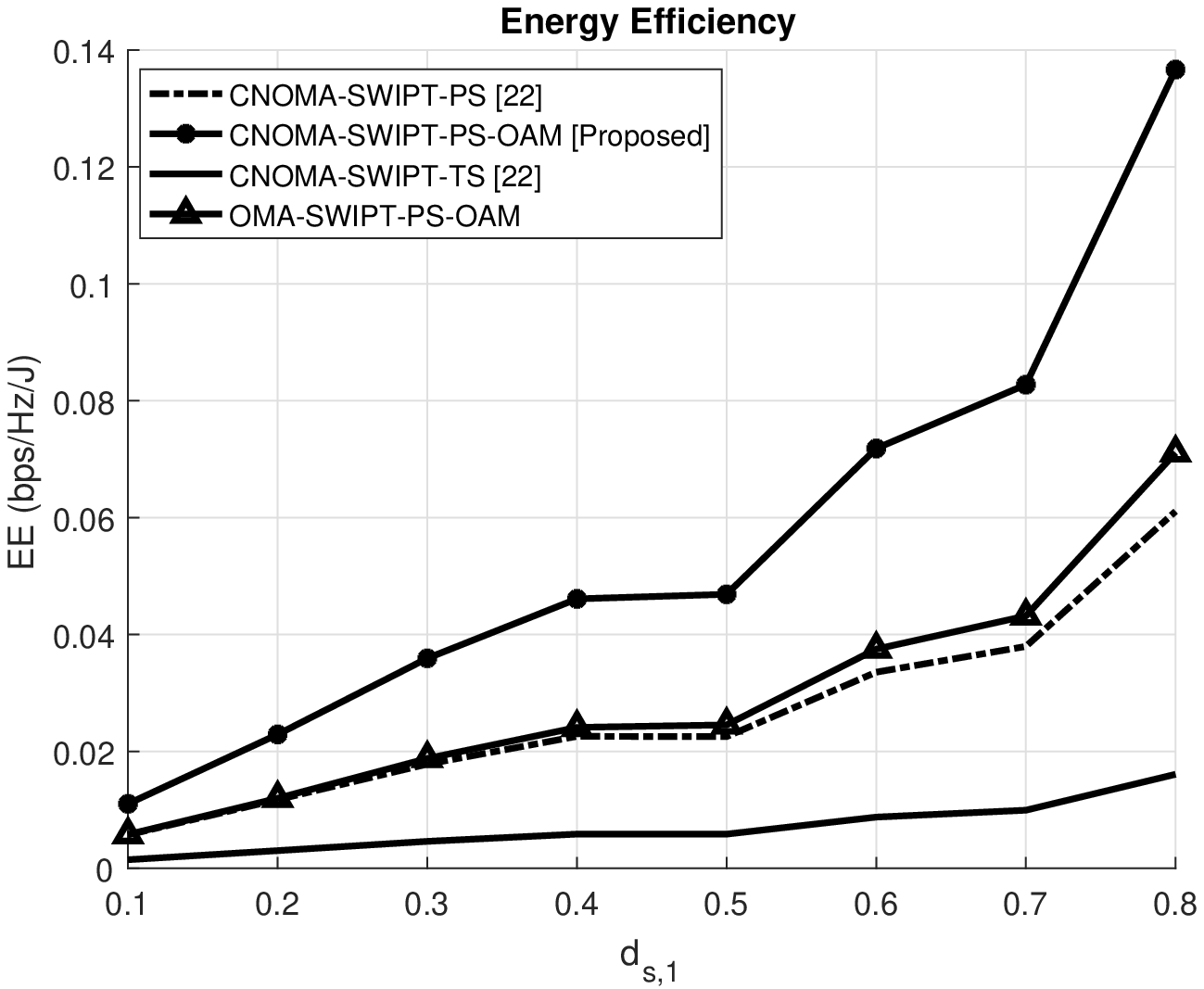}
\caption{EE comparisons with respect to $d_{s,1}$;$p_N=0.4$, $p_F=0.6$, $P=1$, $l_{s,1}=1$, $l_{s,2}=2$, $\delta=0.3$, $\eta=0.7$ and $\rho=15dB$.}
\label{image-myimage}
\end{figure}
\begin{figure}[!t]
\centering
\includegraphics[width=0.8\textwidth]{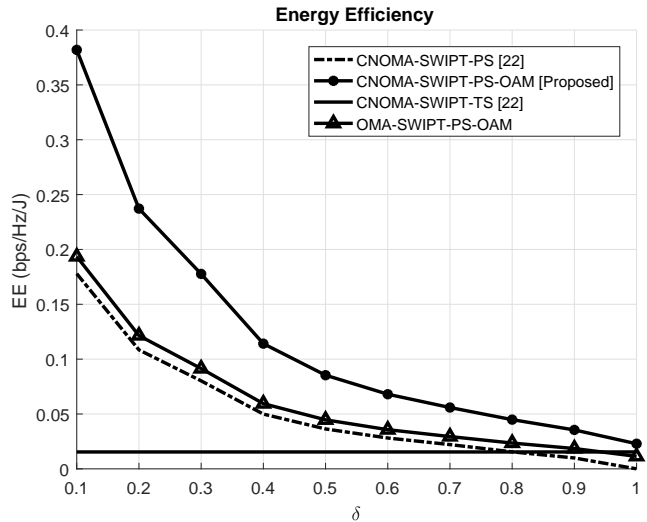}
\caption{EE comparisons with respect to $\delta$; $K_{s,1} = K_{s,2} = 5$, $K_{s,2} = 2$, $p_N=0.4$, $p_F=0.6$, $P=1$, $l_{s,1}=2$, $l_{s,2}=1$, $\eta=0.7$, $d_{s,1}=0.5$, $d_{s,2}=1$ and $\rho=15dB$.}
\label{image-myimage}
\end{figure}
\par
In Figure 9, the EE w.r.t $d_{s,1}$ is plotted for CNOMA-SWIPT-PS-OAM and compared with CNOMA-SWIPT-PS, CNOMA-SWIPT-TS, and OMA-SWIPT-PS-OAM schemes as well. Parameters $p_N=0.4$, $p_F=0.6$, $\eta=0.7$, $\delta=0.3$, and $\rho=15dB$ are set during the simulation purpose. The EE is increasing for all cases in Figure 9. This happens because of the increasing value of $d_{s,1}$ causes a reduction of transmit power for relaying ($P_1$) at CCU due to degradation of channel conditions between BS and CCU. The EE is significantly higher for the proposed CNOMA-SWIPT-PS-OAM scheme than other schemes which is shown in Figure 9. Because according to Figure 6, the SC of the proposed scheme is significantly higher than other schemes due to additional symbol transmission from BS to the users by different OAM modes. Due to utilizing the separate time slot for each transmission except OAM based transmission, the OMA-SWIPT-PS-OAM scheme provides significantly lower EE than the proposed scheme w.r.t $d_{s,1}$ which is illustrated in Figure 9.

\par

In Figure 10, the impact of $\delta$ over EE is plotted for CNOMA-SWIPT-PS-OAM and compared with CNOMA-SWIPT-PS, CNOMA-SWIPT-TS, and OMA-SWIPT-PS-OAM schemes as well. Parameters $p_N=0.4$, $p_F=0.6$, $\eta=0.7$, $d_{s,1}=0.5$, $d_{s,2}=1$, $d_{1,2}=1-d_{s,1}$, and $\rho=15dB$ are set during the simulation purpose. The EE is decreasing for increasing value of $\delta$. According to Figure 7, SC is decreasing for increasing values of $\delta$. Moreover, Figure 7 shows that the proposed CNOMA-SWIPT-PS-OAM provides better SC than other schemes for different values of $\delta$ as well. The additional information transmission by different OAM modes boosts up the SC of the proposed scheme than other schemes significantly. As a result, the EE is higher for the proposed scheme than other schemes which is shown in Figure 10. In addition, there is no impact of $\delta$ on SC of CNMOMA-SWIPT-TS. Because $\delta$ is only used for energy harvesting for PS based SWIPT protocols [22]. So there is no impact of $\delta$ on EE of CNMOMA-SWIPT-TS scheme as well. Due to utilizing the separate time slot for each transmission except OAM based transmission, the OMA-SWIPT-PS-OAM scheme provides significantly lower EE than the proposed scheme w.r.t $\delta$ which is illustrated in Figure 10.

\section{Conclusion}

In this paper, the CNOMA-SWIPT-PS-OAM scheme has been proposed to enhance the capacities and energy efficiency of PS SWIPT based  CNOMA downlink transmission. Moreover, OMA-SWIPT-PS-OAM and other two conventional schemes such as CNOMA-SWIPT-PS and CNOMA-SWIPT-TS are also compared with the proposed scheme for a fair comparison. According to result analysis, it is shown that the proposed scheme provides better performance compared to other conventional schemes in terms of capacities and energy efficiency. In the future, the work can be extended by incorporating amplify and forward relay assisted CNOMA with the proposed scheme.


\begin{acknowledgements}
This work was supported by the National Research Foundation of Korea(NRF) grant funded by the Korea government(MEST) (No. 2019R1A2C1089542)
\end{acknowledgements}



\end{document}